\documentclass[preprint,showpacs]{revtex4}

\usepackage{graphicx}
\usepackage{bm}       
\usepackage{mathptmx} 

\begin{document}

\title{Four-dimensional graphene and chiral fermions}
\author{Michael Creutz}
\affiliation{
Physics Department, Brookhaven National Laboratory\\
Upton, NY 11973, USA
}

\begin{abstract}
{Motivated by the description of the graphene electronic structure in
  terms of the relativistic Dirac equation, a generalization to four
  dimensions yields a strictly local fermion action describing two
  species and possessing an exact chiral symmetry.  This is the
  minimum number of species required by the well known ``no-go''
  theorems.}
\end{abstract}

\maketitle

The structure of graphene, a single layer of graphite consisting of a
hexagonal lattice of carbon atoms, has attracted considerable
attention recently both from the experimental front and the fact that
the low electronic excitations are described by the Dirac equation for
massless fermions \cite{graphene}.  From the point of view of a
particle physicist, this structure has two particularly striking
features.  First, the massless structures are robust for topological
reasons related to chiral symmetry.  Second, it achieves this symmetry
in a manner that involves the minimum number of effective massless
fermions required by the famous ``no-go theorems'' for lattice chiral
symmetry \cite{Nielsen:1980rz,Wilczek:1987kw}.

Given the importance of chiral symmetry in particle physics and the
difficulties with implementing it with a lattice regularization
\cite{Creutz:2000bs}, it is natural to ask whether these properties of
the graphene electronic structure can be extended to four dimensions.
Indeed, this is possible, and provides a remarkable fermion action
with an exact chiral symmetry and manifesting two species of massless
states, the minimal number consistent with chiral symmetry.  This
action is strictly local and thus will be vastly faster in simulations
than either the overlap operator or domain wall fermions, the only
other known ways to have chiral symmetry with only two flavors.

Recently a chiral gauge theory structure on two dimensional graphene
has been proposed \cite{Jackiw:2007rr}.  Given that we do not yet have
a lattice regularization of the standard model, it would be
particularly interesting if this construction could be extended to
the four dimensional lattices presented here.

\begin{figure*}
\centering
\includegraphics[width=3in]{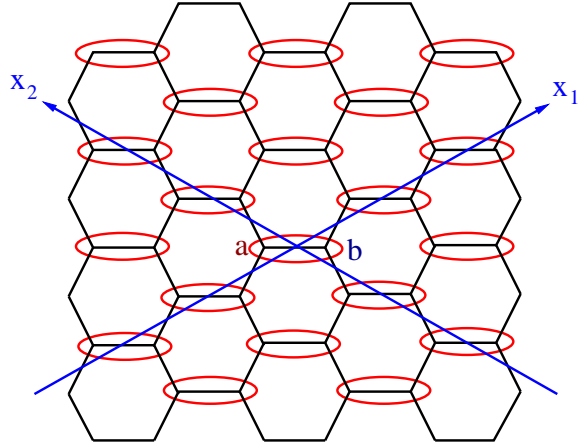}
\caption{\label{graphenefig} Organize the graphene structure into
two-atom ``sites'' involving horizontal bonds as shown by ellipses.
Call the left hand atom of each site type $a$ and the right hand atom
type $b$.  The coordinates of the sites are labeled along the
non-orthogonal $x_1$ and $x_2$ axes.}
\end{figure*}

Although well known, it is useful to review briefly the standard
two-dimensional graphene band structure.  We will closely parallel
this derivation for the four-dimensional case.  Ignored here are all
but the pi orbitals in a tight binding approximation.  We also ignore
the electron spin, each component of which gives an equivalent
structure.  Our electrons hop from neighbor to neighbor around a fixed
underlying hexagonal lattice.  A fortuitous choice of coordinates
makes the problem straightforward to solve.  As sketched in
Fig.~(\ref{graphenefig}), orient a graphene surface with one third of
the bonds horizontal, one third sloping up at 60 degrees, and one
third sloping down.  It is then convenient to collapse the atoms at
the opposite ends of each horizontal bond together and call this unit
a lattice ``site,'' as enclosed in ellipses in the figure.  For each
site, let $a^\dagger$ denote the creation operator for an electron on
the left atom, and correspondingly let $b^\dagger$ create an electron
on the right atom.  The commutation relations are the usual
\begin{equation}
[a_{x_1,x_2},a^\dagger_{x_1^\prime,x_2^\prime}]_+
=[b_{x_1,x_2},b^\dagger_{x_1^\prime,x_2^\prime}]_+
=\delta_{x_1,x_1^\prime}\delta_{x_2,x_2^\prime}
\end{equation}
with the $a$ type operators anti-commuting with the $b$'s.  Finally, it
is useful to label the sites using a non-orthogonal coordinate system
with axes $x_1$ sloping up at 30 degrees intersecting the
corresponding sites, and similarly $x_2$ sloping down at 30 degrees.
All of this is illustrated in Fig.~(\ref{graphenefig}).

With these conventions, the Hamiltonian of interest involves only
nearest neighbor hoppings between $a$ and $b$ type sites
\begin{equation}
\matrix{
H=K\sum_{x_1,x_2} \big(
a_{x_1,x_2}^\dagger b_{x_1,x_2}
+b_{x_1,x_2}^\dagger a_{x_1,x_2}\cr
+a_{x_1+1,x_2}^\dagger b_{x_1,x_2}
+b_{x_1-1,x_2}^\dagger a_{x_1,x_2}\cr
+a_{x_1,x_2-1}^\dagger b_{x_1,x_2}
+b_{x_1,x_2+1}^\dagger a_{x_1,x_2}\big).\cr
}
\end{equation}
Here $K$ is the basic hopping parameter.  The phase of $K$ is a
convention; here I consider positive real hopping.  To diagonalize
this Hamiltonian go to momentum space
\begin{equation}
a_{x_1,x_2}=\int_{-\pi}^\pi {dp_1\over 2\pi}\ {dp_2\over 2\pi}
\ e^{ip_1 x_1}\ e^{ip_2x_2}
\ \tilde a_{p_1,p_2}.
\end{equation}
This brings the Hamiltonian to the simple form
\begin{equation}
\matrix{
H=K\int_{-\pi}^\pi {dp_1\over 2\pi}\  {dp_2\over 2\pi}
&\tilde a^\dagger_{p_1,p_2}\tilde b_{p_1,p_2}
(1+e^{- ip_1}+e^{+ i p_2})\cr
&+\tilde b^\dagger_{p_1,p_2}\tilde a_{p_1,p_2}
(1+e^{+ip_1}+e^{- i p_2}).\cr
}
\end{equation}
The problem reduces to diagonalizing a two by two matrix of form
\begin{equation}
\label{hmatrix}
H(p_1,p_2)=K\pmatrix{
0 & z\cr
z^* & 0\cr
}
\end{equation}
with
\begin{equation}
\label{threeterms}
z= 1+e^{- ip_1}+e^{+i p_2}.
\end{equation}
The energy eigenvalues are 
\begin{equation}
E(p_1,p_2)=\pm K|z|.
\end{equation}
From Eq.~(\ref{threeterms}) it is easy to see that the energy vanishes
only at two points, $p_1=p_2=\pm 2\pi/3$.

These are known as ``Fermi points'' \cite{volovik} and their robustness can
be seen by considering contours of constant energy.  These are closed
curves of constant $|z|$ in $p_1,p_2$ space .  The important point is
that for such a contour near one of the zero energy solutions, the
phase of $z$ wraps non-trivially around a circle.  This
non-contractable mapping indicates that on reducing the energy and
shrinking the curve to a point, the magnitude of the energy at this
point must vanish.  This is the mechanism that prevents a band gap
from opening in the spectrum.

This robustness is associated with a chiral symmetry.  Because the
hoppings are always between $a$ and $b$ type sites, we can change the
sign of the energy by taking $b\rightarrow -b$.  This is equivalent to
the statement that $\sigma_3$ anti-commutes with the Hamiltonian.  For
the four-dimensional generalization, this will become the
anti-commutation of $\gamma_5$ with the Dirac operator.

We wish to extend this formalism to the four-dimensional case.  We
want an operator $D$ to insert into the Euclidian path integral via
the fermion action $\overline\psi D\psi$.  For low energy excitations
this operator should reduce to two massless Dirac fermions, and this
reduction should be robust due to a chiral symmetry.  Essentially all
Dirac operators used in practice for lattice gauge theory satisfy what
is called a ``$\gamma_5$ Hermiticity'' condition
\begin{equation} 
 \gamma_5 D\gamma_5=D^\dagger
\end{equation} 
where $\gamma_5$ is the usual four by Dirac matrix.  It is Hermitean
and squares to the unit matrix.  A specific realization of this and
the other Dirac matrices will appear below.  Using this we can
construct a Hermitean ``Hamiltonian''
\begin{equation}
H=\gamma_5 D
\end{equation}
with which we will parallel the two dimensional discussion.  It is
important to remember that this is not the Hamiltonian of the three
dimensional quantum system, but a convenient operator for leading us
back to $D$.  In four-dimensional space, the analog of the curves of
constant ``energy'' are three dimensional manifolds.  To maintain a
topological argument in analogy to the two dimensional case, we want
to consider the situation where these surfaces wrap non-trivially
around a three sphere, an $S_3$.  For this purpose it is quite natural
to maintain the form of Eq.~(\ref{hmatrix}), but extend $z$ to two by
two matrices in a quaternionic space.  That is, take
\begin{equation}
z=a_0+i\vec a\cdot\sigma
\end{equation}
with $a_\mu$ a real four vector and $\vec\sigma$ denotes the
traditional Pauli matrices.  
With $z$ being a two by two matrix inserted into the two by two matrix
of Eq.~(\ref{hmatrix}), we wind up with a four by four matrix, the
same dimension as used in the usual Dirac equation.
As before, vanishing energy states occur when $z$ vanishes, which now
corresponds to $a_\mu$ vanishing as a four vector.  The goal is to
construct our Hamiltonian so that that constant energy surfaces that
wrap around zero energy points within the Brillouin zone will involve
a non-trivial mapping in the quaternionic space.  Because of the
periodicity of the Brillouin zone, these zero energy points must
appear in pairs so that the overall wrapping will vanish.  Indeed,
this is the famous no-go theorem \cite{Nielsen:1980rz,Wilczek:1987kw}.

We want a construction giving precisely one and only one such pair.
We also want to involve only local couplings, i.e. with only simple
trigonometric functions of the momenta appearing in the dispersion
relations.  Because of the robustness of the zeros, if such a
construction exists, it is clearly not unique.  To find one such
solution, start with a regular four-dimensional lattice and perform a
Fourier transform.  Now there will be four momentum variables
$p_1,p_2,p_3,p_4$, all ranging from $-\pi$ to $\pi$.  A convenient
form to explore is
\begin{equation}
\label{z}
\matrix{
z=&B\big(4C-\cos(p_1)-\cos(p_2)-\cos(p_3)-\cos(p_4)\big)\cr
&+i\sigma_x \big(\sin(p_1)+\sin(p_2)-\sin(p_3)-\sin(p_4)\big)\cr
&+i\sigma_y \big(\sin(p_1)-\sin(p_2)-\sin(p_3)+\sin(p_4)\big)\cr
&+i\sigma_z \big(\sin(p_1)-\sin(p_2)+\sin(p_3)-\sin(p_4)\big).\cr
}
\end{equation}
Here $B$ and $C$ are parameters whose values will be discussed later.

For zero energy states we need $z$ to vanish.  This gives four
equations corresponding to the coefficients of 1 and each $\sigma_i$
vanishing.  From the coefficients of the Pauli matrices imply the
sines of all the momentum components must be equal.  Picking $p_1$
arbitrarily, each other $p_\mu$ must either equal $p_1$ or $\pi-p_1$.
Now turning to the constant part of $z$, we have
\begin{equation}
\cos(p_1)+\cos(p_2)+\cos(p_3)+\cos(p_4)
=4C.
\end{equation}
Since the cosine function is bounded by unity, we clearly must take
$C<1$ to have any solutions. To resolve the $p_i\leftrightarrow
\pi-p_i$ ambiguity it is convenient to ask that $\cos(p_i)$ be
positive.  Imposing the constraint $C>1/2$ ensures this.  We will
later discuss some interesting specific choices for $C$.

So with these constraints on the constant $C$ there are exactly two
zeros of energy in the Brillouin zone.  These occur when all
components of $p$ are equal and satisfy $\cos(p)=C$.  The two
solutions differ in the sign of $p$.  Picking the positive sign for
convenience, it is useful to expand about the zero
\begin{equation}
p_\mu=\tilde p+q_\mu
\end{equation}
with $\cos(\tilde p)=C$ and $\tilde p>0$.  Defining $S=\sin(\tilde p)
=\sqrt{1-C^2},$ we have
\begin{equation}
\matrix{
\cos(p_\mu)=C\cos(q_\mu)-S\sin(q_\mu)=C-Sq_\mu+O(q^2)\cr
\sin(p_\mu)=S\cos(q_\mu)+C\sin(q_\mu)=S+Cq_\mu+O(q^2).\cr
}
\end{equation}
Inserting all this into our quaternion
\begin{equation}
\matrix{
z=BS(q_1+q_2+q_3+q_4)\cr
+iC\sigma_x (q_1+q_2-q_3-q_4)\cr
+iC\sigma_y (q_1-q_2-q_3+q_4)\cr
+iC\sigma_z (q_1-q_2+q_3-q_4)&+O(q^2).\cr
}
\end{equation}
This two by two matrix is to be inserted into the analogue of
Eq.~(\ref{hmatrix}) to otain a four by four matrix.  At this point we
introduce a convention for the Dirac gamma matrices
\begin{equation}
\matrix{
\vec\gamma=\sigma_x\otimes\vec\sigma\cr
\gamma_4=\sigma_y\otimes 1\cr
\gamma_5=\sigma_z\otimes 1=\gamma_1\gamma_2\gamma_3\gamma_4\cr
}
\end{equation}
The direct product notation here is defined so that $\gamma_5$ is
diagonal with $-1$ in the last two places.  With these conventions our
Euclidean Dirac operator takes the form
\begin{equation}\matrix{
D=
C(q_1+q_2-q_3-q_4)i\gamma_1\cr
+C(q_1-q_2-q_3+q_4)i\gamma_2\cr
+C(q_1-q_2+q_3-q_4)i\gamma_3\cr
+BS(q_1+q_2+q_3+q_4)i\gamma_4+O(q^2).\cr
}
\end{equation}
This reproduces the desired massless Dirac equation if we identify new
momenta
\begin{equation}\matrix{
k_1=C(q_1+q_2-q_3-q_4)\cr
k_2=C(q_1-q_2-q_3+q_4)\cr
k_3=C(q_1-q_2+q_3-q_4)\cr
k_4=BS(q_1+q_2+q_3+q_4).\cr
}
\end{equation}
Proper Lorentz invariance requires symmetry between the $k$'s.  This
implies that the original lattice, as generated by translations using
the $q$'s, will in general be distorted from simple hyper-cubic.

The physical angles between the original lattice directions are easily
determined.  For example, a step along the $q_1$ axis represents a
shift in the $k_\mu$ basis by $(C,C,C,BS)$ while going along the $q_2$
axis gives $(C,-C,-C,BS)$.  These vectors are at the angle
\begin{equation}
\cos(\theta)={B^2S^2-C^2 \over B^2S^2 +3C^2}
\end{equation}
to each other.  

Note that the original axes can be made orthogonal by setting
$B=C/S=\cot(\tilde p)$.  With such a choice, gauging the theory is
straightforward.  Because the starting links are orthogonal, we can
use the usual Wilson gauge action with group elements on links
interacting with the simple plaquette action.  Borici
\cite{Borici:2007kz} has recently suggested using $B=1$ and
$S=C=1/\sqrt 2$ as a particularly simple case with orthogonal axes.

Another especially interesting choice for the parameters $B$ and $C$
gives a closer analogy with graphene.  Imagine the fermion fields on
one site to actually be spread along a new bond connecting two atoms
in the $k_4$ direction, similar to the construction for graphene
indicated in Fig.~(\ref{graphenefig}).  This structure would be
particularly symmetric if the angles between all five bonds attached
to an atom were equal in length and distributed at a common angle to
each other.

To construct this structure it is simplest to look in momentum space
continued beyond the first Brillouin zone.  The zeros of the energy
then form a lattice on which we wish to impose the above symmetry.
The idea is to consider one zero and ask that the five closest ones
are equally distant in the physically scaled momentum units and at
angles from each other satisfying $\cos(\theta)=-1/4$.  More
precisely, ask that the physical four vector between the two zeros at
$p_\mu=\tilde p$ and $p_\mu=-\tilde p$ be of the same physical length
as the vector from the first of these to the repetition of the second
when any one of the four components of $p$ is increased by $2\pi$.  We
obtain a set of five four vectors, which should be symmetrically
distributed in four space.  The conditions on lengths and angles
then determines the parameters as
\begin{equation}
\label{symmetric}
\matrix{ \tilde p={\pi\over 5}\cr
C=\cos(\pi/5)\cr
B=\sqrt 5 \cot(\pi/5).\cr
}
\end{equation}

This lattice of zeros has an appealing intuitive geometric
interpretation in terms of bonds along one direction, analogous to the
horizontal bonds in Fig.~(\ref{graphenefig}), splitting at a site into
four bonds going off symmetrically into four-space.  These then join a
repetition of this structure with new horizontal bonds displaced in
the various directions.  While in two dimensional graphene each carbon
is coupled symmetrically to three neighbors, here each atom is
directly coupled to 5 others.  The entire lattice is then built up of
hexagonal chairs with an inter-bond angle of ${\rm
acos}(-1/4)=104.4775\ldots$ degrees.  Note that the diamond lattice in
three dimensions represents an intermediate case, where one bond
splits into three giving each atom a tetrahedral environment.

This lattice has a 120 element discrete symmetry group under
permutations of the five neighbors.  The specific action chosen here,
however, selects the $k_4$ direction as special from the way it
appears in the site diagonal term.  Because of this, the full action
is not exactly invariant under the full 120 element group, but only a
24 element subgroup of rotations and inversions that leave invariant
the positive diagonal of the initial hypercubic structure defined by
the $q$ axes.  This tetrahedral symmetry, which is that of the three
dimensional diamond lattice, applies for all valid values of the
parameters $B$ and $C$ and includes 12 parity odd operations.
Considering the time axis to run along this positive diagonal, there
is in addition a symmetry under reflections about a spacelike surface
perpendicular to this time axis accompanied by a rotation to bring the
bonds back into alignment.  These symmetries will eliminate many
possible lattice artifacts and constrain operator mixing in
renormalization studies \cite{Bedaque:2008xs}.  The main remaining
issue is that the special treatment of time still allows a
renormalization of the speed of light.  This can be compensated by
adding to the gauge action terms involving six link chairs orthogonal
to the time direction.  While such terms are generally necessary, for
the hyper-cubic values $B=C/S$ they should go to zero in the continuum
limit.  Indeed, these are all issues of order the lattice spacing
squared, as discussed in \cite{Cichy:2008gk}.

From the standpoint of computational efficiency, it does not appear to
matter much what values of the parameters $B$ and $C$ are chosen,
within the constraint $1/2<C<1$.  Near the limits of this range one
should expect lattice artifacts to increase.  The hyper-cubic values
$B=C/S$ are presumably closest to conventional lattice gauge ideas and
simplest to gauge, with only a traditional plaquette action required.
On the other hand, the values in Eq.~(\ref{symmetric}) may have
smaller lattice artifacts due to the high symmetry.  

Returning from reciprocal to position space, the fermionic action
involves several terms.  First from the collapsed $k_4$ bond there is
a site diagonal term $4iBC\overline\psi \gamma_4\psi$.  Then for a
forward step in the various directions we pick up a factor of the
hopping parameter $K$ multiplied by different combinations of gamma
matrices, as listed here:
\begin{equation}
\matrix{
&\hbox {for a hop in direction 1:}&
 +\gamma_1+\gamma_2+\gamma_3-iB\gamma_4\cr
&\hbox {for a hop in direction 2:}&
 +\gamma_1-\gamma_2-\gamma_3-iB\gamma_4\cr
&\hbox {for a hop in direction 3:}&
 -\gamma_1-\gamma_2+\gamma_3-iB\gamma_4\cr
&\hbox {for a hop in direction 4:}&
 -\gamma_1+\gamma_2-\gamma_3-iB\gamma_4\cr
}
\end{equation}
Keeping the operator $D$ anti-Hermitean, the reverse hops involve minus
the conjugate of these factors.  Note the factor of $i$ in front of
$\gamma_{4}$ which is absent for the $\gamma_{1-3}$ terms.  This
twisting of the phase gives rise to the required factors of $\sin$ or
$\cos$ in Eq.~(\ref{z}).  This action is only marginally more complicated
than that of naive fermions; so, it should be easy to insert into
simulations.

Chiral symmetry is manifested in the exact anti-commutation of
$\gamma_5$ with $D$.  This is actually a flavored chiral symmetry
since the expansion about the negative solution for $p$ flips the sign
of the gamma matrix associated with $k_0$.  Note that as with naive
fermions, $D$ is purely anti-Hermitean.  The chiral symmetry can
easily be broken with the addition of a term proportional to
$\gamma_5$ to $H$ that splits the degeneracy of the type $a$ and type
$b$ sites.  This gives each physical fermion a common mass.  Because
the chiral symmetry remains exact on gauging, an additive
renormalization for the physical fermion mass is forbidden, unlike
with Wilson fermions.  The further addition of a Wilson type mass term
would enable splitting the degeneracy of the two species.  Note that
if we wish to extend this formalism to more flavors, the no-go theorem
restricts us to an even number of species.  The best we can do for
three flavors is to start with four and, using a chiral symmetry
breaking operator, make one of the flavors heavier.  Such a term will
unfortunately introduce the possibility of an additive mass
renormalization for the third flavor.

It is perhaps interesting to compare this formalism with staggered
fermions \cite{Kogut:1974ag,Susskind:1976jm,Sharatchandra:1981si}.
The latter also have an exact chiral symmetry and the appearance of
several degenerate flavors (sometimes called tastes to distinguish
them from independent lattice fields).  In both formalisms the chiral
symmetry is flavored, with the different species rotating differently
under a chiral rotation.  Also, in both cases mixing between the
flavors will result in the zero modes associated with topologically
non-trivial gauge fields to no longer be exact at finite lattice
spacing \cite{Karsten:1980wd,Smit:1986fn}.  However the usual
staggered approach has the doublers appearing in multiples of four,
and thus is further from the situation desired for the physical case
with only the light up and down quarks.  Staggered fermions do have
only one component of the field per site, yielding potentially faster
simulations at the expense of an intricate non-local formalism in
constructing composite fields.

\section*{Acknowledgements}
This manuscript has been authored under contract number
DE-AC02-98CH10886 with the U.S.~Department of Energy.  Accordingly,
the U.S. Government retains a non-exclusive, royalty-free license to
publish or reproduce the published form of this contribution, or allow
others to do so, for U.S.~Government purposes.

\end{document}